\documentclass[conference]{IEEEtran}

\usepackage{amsmath,amsfonts, times,booktabs,tabularx,amsmath,graphicx,url,cite,balance,amssymb, amsthm, array, multirow, dblfloatfix, float, algorithmic, textcomp, xcolor}
\def\BibTeX{{\rm B\kern-.05em{\sc i\kern-.025em b}\kern-.08em
    T\kern-.1667em\lower.7ex\hbox{E}\kern-.125emX}}
\begin{document}

\title{Leveraging Audio-Tagging Assisted Sound Event Detection using Weakified Strong Labels and Frequency Dynamic Convolutions\\}

\author{\IEEEauthorblockN{Tanmay Khandelwal\IEEEauthorrefmark{1},
 Rohan Kumar Das\IEEEauthorrefmark{1}, Andrew Koh\IEEEauthorrefmark{2} and
Eng Siong Chng\IEEEauthorrefmark{2}}
\IEEEauthorblockA{\IEEEauthorrefmark{1}Fortemedia Singapore, Singapore \IEEEauthorrefmark{2}Nanyang Technological University, Singapore \\
Email: f20170106p@alumni.bits-pilani.ac.in,
rohankd@fortemedia.com,
andr0081@e.ntu.edu.sg,
aseschng@ntu.edu.sg
}}
\maketitle

\begin{abstract}
Jointly learning from a small labeled set and a larger unlabeled set is an active research topic under semi-supervised learning (SSL). In this paper, we propose a novel SSL method based on a two-stage framework for leveraging a large unlabeled in-domain set. Stage-1 of our proposed framework focuses on audio-tagging (AT), which assists the sound event detection (SED) system in Stage-2. The AT system is trained utilizing a strongly labeled set converted into weak predictions referred to as weakified set, a weakly labeled set, and an unlabeled set. This AT system then infers on the unlabeled set to generate reliable pseudo-weak labels, which are used with the strongly and weakly labeled set to train a frequency dynamic convolutional recurrent neural network-based SED system at Stage-2 in a supervised manner. Our system outperforms the baseline by 45.5\% in terms of polyphonic sound detection score on the DESED real validation set.
\end{abstract}


\section{Introduction}
\label{sec:intro}
Sound aids us in perceiving environmental changes and comprehending our surroundings. Humans have an in-built system for detecting and categorizing sound events in our   
 various environments. The SED applications include audio surveillance in a variety of environments, such as smart-homes~\cite{home, babycrywork} and cities~\cite{smarthome, smarthome2}. 

The models developed for SED require strongly labeled data to accurately predict the temporal onset and offset. The manual annotation process for generating strong labels is expensive and time-consuming, and the annotations vary greatly due to the subjective judgment of the annotators. On the other hand, annotating the entire clip with audio labels to generate weak labels is much easier. Furthermore, collecting unlabeled datasets in-domain is equally simple. To leverage this readily available unlabeled set with a small amount of labeled set, several previous works have employed semi-supervised learning (SSL) techniques. The authors of~\cite{meanteacher} used mean-teacher (MT) learning method that employs exponential moving average, whereas an unsupervised data-augmentation is used in~\cite{uda}, which enforces the model to be consistent with respect to the noise added using data-augmentation (DA) techniques. In~\cite{skunit}, the authors used interpolation consistency training (ICT) and shift consistency training, whereas in~\cite{pseudo}, they self-trained to produce pseudo-labels and train on them.

To use the unlabeled set for supervised learning, the model generates pseudo-labels~\cite{stage, pseudoweak, noisy}. The pseudo-labeling process is similar to entropy minimization~\cite{entropy} and helps in cases where it can recover the cluster structure among the various classes~\cite{pseudo-labeling}. It requires a sufficient number of labeled points to effectively learn the differentiation between the clusters. The labels for pseudo-labels are determined by the confidence threshold. The clip-wise labels above the confidence threshold are used as true labels for clips~\cite{noisy} in the typical supervised loss function. The model is then trained using labeled and unlabeled sets simultaneously. In addition to SSL techniques, past works have employed various DA techniques like SpecAugment~\cite{specaug}, time-shift~\cite{stage}, pitch-shift~\cite{pitch}, and mixup~\cite{mixup} to increase diversity and reduce overfitting. 

The detection and classification of acoustic scenes and events (DCASE) 2022 Task 4 focuses on SED-based SSL to utilize labeled and unlabeled data. We make the following contributions in this work to effectively exploit the unlabeled in-domain set provided in DCASE 2022 Task 4 by generating pseudo-labels:
\begin{itemize}
  \item Proposal of a two-stage framework~\cite{Khandelwal2022} that performs audio-tagging (AT) at the first stage to estimate reliable pseudo-labels on unlabeled data used to train the SED system at the second stage in a supervised manner. 
  \item A novel weak training strategy to create weakified labels, where the strong labels are converted to weak predictions. The objective is to supply more weak labels for Stage-1 system training to lessen the model's inclination to predict inactive frames~\cite{filteraug} when trained with strong labels. 
  \item Utilize pre-trained audio neural networks (PANNs) for Stage-1 to further improve the reliability of the pseudo-weak labels used to train the Stage-2 system. 
\end{itemize}
We used several DA techniques, pooling functions, and adaptive post-processing to improve the robustness of the developed systems, evaluate the proposed method, and make fair comparisons with other state-of-the-art methods.

\section{Sound Event Detection System}
\label{sec:architecture}
In this section, we briefly review the baseline system and the proposed, two-stage framework for sound event detection.
\subsection{Baseline}
\label{ssec:baseline}
The baseline~\cite{baseline} architecture is a combination of convolutional neural network (CNN) and recurrent neural network (RNN) called convolutional recurrent neural network (CRNN), as depicted in Fig.~\ref{fig:fdy} (a). The CNN part is made up of 7-blocks, each with 16, 32, 64, 128, 128, 128, and 128 filters, respectively. It has a kernel size of $3\times3$ and an average-pooling of [2, 2], [2, 2], [2, 1], [2, 1], [2, 1], [2, 1], [2, 1] per layer. The RNN is composed of two layers of 128 bidirectional gated recurrent units (Bi-GRU)~\cite{gru}. The RNN block is followed by an attention pooling layer, which is a multiplication of a linear layer with softmax activations and a linear layer with sigmoid activations. The baseline employs the MT~\cite{meanteacher} strategy, which is a hybrid of two models: the student model and the teacher model (both having the same architecture). The student model is the final model used for inference, whereas the teacher model is designed to help the student model during training. Its weights are an exponential moving average of the student model's weights.

\subsection{Proposed two-stage framework}
\label{ssec:two-stage}
The objective of any SSL algorithm is to utilize labeled and unlabeled data to learn the underlying structure of the dataset effectively. The small amount of labeled data helps the model learn discrete or non-overlapping clusters for different labels. The cluster assumption [7] states that close points have the same class and points in different classes are more widely separated, therefore true decision boundaries flow through low-density input space. As training progresses, these clusters improve their cluster boundary, improving model predictions on the unlabeled in-domain set. To mitigate the problem of a small labeled set and to utilize the unlabeled set by learning the discrete clusters for each class, we propose a two-stage framework~\cite{Khandelwal2022}, shown in Fig.~\ref{fig:stage}. Stage-1 utilizes the proposed weak training method to focus on AT, and Stage-2 then utilizes the reliable pseudo-labels generated from Stage-1 to have an improved SED performance. Furthermore, each stage makes use of MT adopted from the baseline. In addition to MT, we use another method used for SSL, called ICT~\cite{ict}, in both stages of the two-stage framework. The ICT substitutes all input samples with interpolated samples, helping the model to improve the generalization ability. A detailed description of the models used in each stage is given in the following subsections.

\subsubsection{Stage-1}
\label{ssec:stage1}
\begin{figure}[t]
\centering  
\includegraphics[width=1\columnwidth]{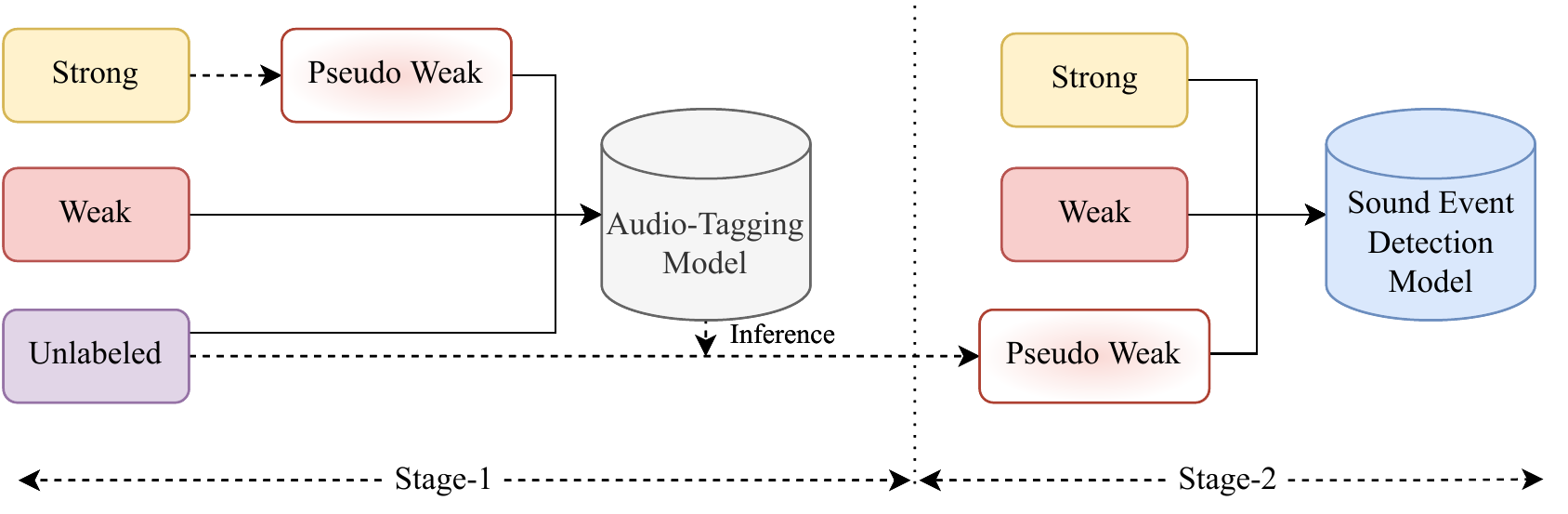}
\caption{Proposed two-stage learning setup, with Stage-1 focusing on AT and Stage-2 focusing on SED.}
\label{fig:stage}
\vspace*{-4pt}
\end{figure}

In order to have an effective AT in Stage-1, \cite{stage} showed the importance of deeper neural network models compared to the baseline CRNN. As feature extractor, we used CNN-14-based PANNs~\cite{panns} with 118M parameters for pre-trained embeddings. The parameters of the PANNs-based embeddings are unfrozen and trained. The 14-layer CNN feature extractor consists of 6 convolutional blocks. Each convolutional block consists of 2 convolutional layers with a kernel size of $3\times3$. In addition, each convolutional layer is followed by batch normalization and rectified linear unit~\cite{relu} non-linearity to stabilize the training. Average pooling~\cite{avgpool} of $2\times2$ is applied to each convolutional block for down-sampling. The feature extractor is followed by 2-layers of Bi-GRU with 1024 hidden units. For frame-level predictions, the RNN output is multiplied by a dense layer with sigmoid activation, and for clip-level predictions, the linear layer is multiplied by a dense layer with softmax activation. Based on previous work~\cite{filteraug} that uses only the weakly labeled data, we suggest a weak training strategy to improve Stage-1 AT systems. We converted the strongly labeled set into a weakly labeled set by removing the onset and offset and keeping the event labels, which we refer to as {\emph {weakified labels}}. Then we trained the AT system using the weakified labels, weakly labeled set, and unlabeled set as illustrated in Fig.~\ref{fig:stage}.

\begin{figure}[t]
\centering  
\includegraphics[width=0.8\columnwidth]{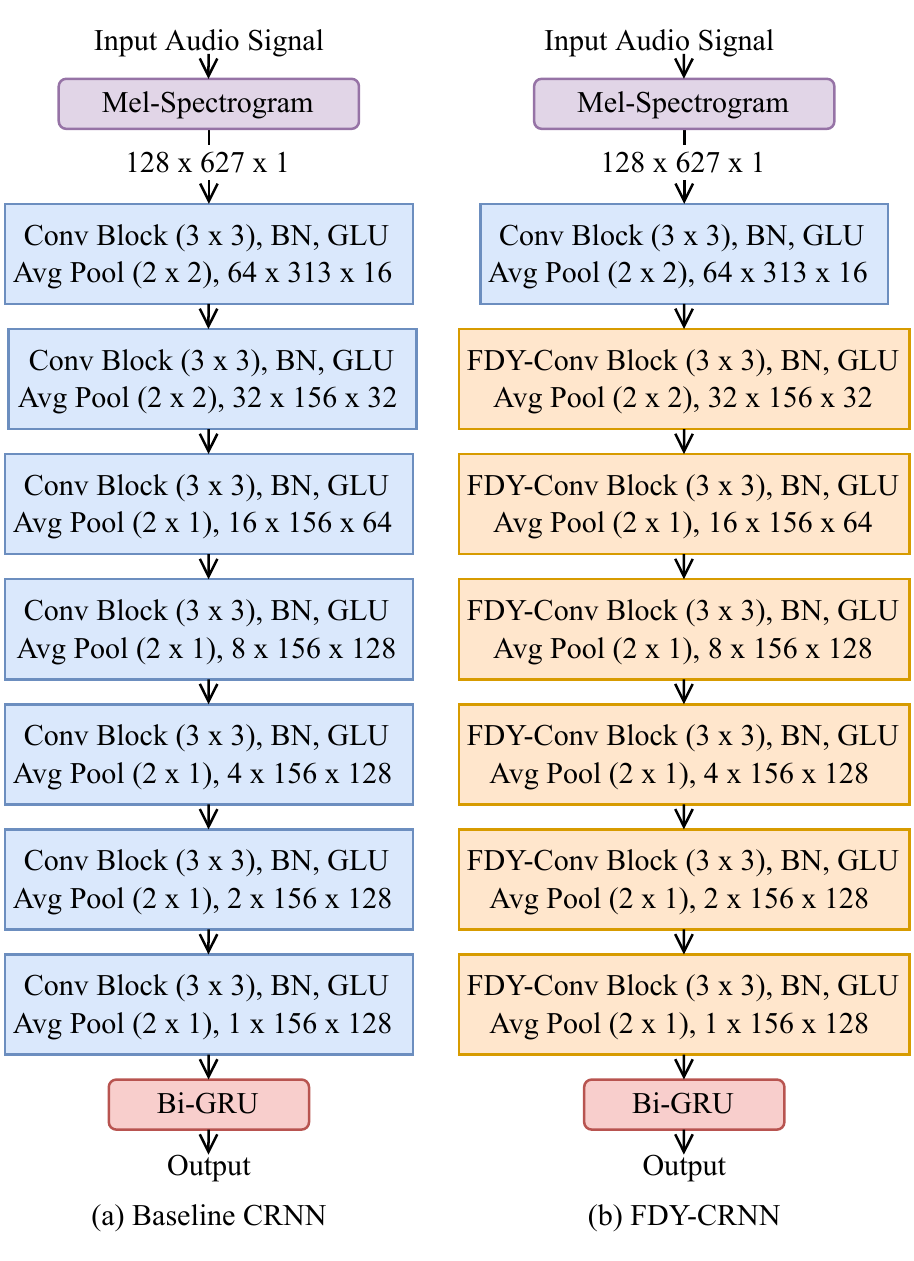}
\caption{Architecture of (a) CRNN (Baseline) (b) FDY-CRNN.}
\label{fig:fdy}
\vspace*{-3pt}
\end{figure}

\subsubsection{Stage-2}
\label{ssec:stage2}
In this work, we used the AT (Stage-1) based system to make predictions on the unlabeled set to use them as pseudo-weak labels in Stage-2 training, as shown in Fig.~\ref{fig:stage}. We believe this way, we can generate reliable pseudo-labels, which can help the SED model at Stage-2. The baseline CRNN's standard 2D convolutional block enforces translation equivariance on sound events along both the time and frequency axes, despite the fact that frequency is not a shift-invariant dimension. To focus on frequency-dependent patterns and to further improve the SED performance, we employed frequency dynamic (FDY)-convolutions proposed in~\cite{fdy} as it applies frequency adaptive kernels to enforce frequency dependency on 2D convolution. We replaced the baseline's standard 2D convolutional blocks with FDY-convolutional blocks, which have the same number of layers and feature maps as that of the baseline, as illustrated in Fig.~\ref{fig:fdy} (b). Then it was trained on a pseudo-weakly labeled set, in addition to the strongly labeled set and the weakly labeled set, in a supervised manner.

\section{ADDITIONAL METHODS}
\label{sec:methods}
\subsection{Pooling function}
\label{ssec:pooling}
Motivated from a prior work~\cite{pooling}, we used exponential softmax to replace the attention pooling used in the baseline. The exponential softmax function assigns a weight of $\exp(y_i)$ to the frame-level probability $y_i$ as given below:
\begin{equation}
   y = \frac{\sum_{i}y_i \exp(y_i)}{\sum_{i}\exp(y_i)}
\end{equation}
where $y_i$ is the predicted probability of an event occurring in the $i^{\text{th}}$ frame. This implies that, with a higher prediction probability, the higher the exponential weight is assigned to the frame-level probability. Hence, it is better under the stringent evaluation criteria for the correctness of the category.

\subsection{Asymmetric focal loss (AFL)}
\label{ssec:afl}
AFL~\cite{afl} function is used to control the training weight depending on the ease and difficulty of the model training. The AFL function for each ${k}^{\text{th}}$ data point with target sound event as $y_k$ and predicted sound event as $p_k$ is given below:
\begin{equation}
l_{AFL}(p, y) = \sum_{n=1}^{K}[(1-p_k)^\gamma y_k\ln p_k + (p_k)^\zeta (1-y_k)\ln(1-p_k)]
\end{equation}
where the parameters $\gamma$ and $\zeta$ are the weighing hyperparameters given as the input to the function that controls the weight of active and inactivate frames.

\subsection{Data-augmentation (DA)}
\label{ssec:dataaug}
We used several DA techniques during the training in both stages, such as time-masking~\cite{specaug}, frame-shifting~\cite{filteraug}, mixup~\cite{mixup}, addition of Gaussian noise and filter augmentation~\cite{filteraug}. Time-masking masks the sequential time frames (which means replacing the elements by zeros or other values), whereas frame-shifting shifts the features and labels along the time axis. Again, mixup randomly mixes selected samples with a mixing parameter, helping in linear interpolation to improve the robustness of the model. In addition, filter augmentation, which uses varying weights on random frequency regions, has been shown to significantly improve SED performance.

\subsection{Adaptive post-processing}
\label{ssec:postprocessing}
We used adaptive post-processing~\cite{postprocessing} in all trials, where the median filter window sizes $(Win)$ are different for each event category $c$ based on the real-life event lengths as shown below: 
\begin{equation}
   Win_c = duration_c \times \beta_c
\end{equation}
For several event categories with high duration variance, we used $duration_c$ as the median duration. Here, we used $\beta_c = \frac{1}{3}$ and then slightly adjusted the window sizes on the validation set. 

\section{EXPERIMENTAL SETUP}
\label{sec:setup}
The next subsections outline our experimental setup to demonstrate the efficacy of the proposed methods.

\subsection{Dataset}
\label{ssec:dataset}
The DCASE 2022 Task 4 dataset used in this work is composed of 10 seconds audio clips to simulate a domestic environment. The development training set is divided into 3 major subsets:
\begin{itemize}
\item 1,578 real recordings with weak annotations.
\item 14,412 real recordings, unlabeled in-domain training set
\item 10,000 synthetic recordings with strong annotations~\cite{impact}.
\item An additional subset from the recently released strongly labeled AudioSet~\cite{audioset} subset of 3,470 real recordings with strong annotations is released as external data.
\end{itemize}
The development validation set has 1,168 real recordings with strong annotations, and the public evaluation ("YouTube") set has 692 YouTube clips.

\subsection{Pre-processing}
\label{ssec:preprocessing}
The audio clips are re-sampled at 16 kHz to a mono channel. Then, log-mel spectrograms are produced using mel-filters in the frequency domain from 0 to 8 kHz with a window size of 2048 samples and a hop size of 256 samples. Stage-1 employed 64 mel-filters, while Stage-2 used 128 mel-filters. The clips with a duration of less than 10 seconds are padded with silence.

\subsection{Training process}
\label{ssec:training}
The batch size for all the experiments is 48 (1/4 strong set, 1/4 weak set, 1/2 unlabeled set). We employed Adam optimizer~\cite{adam} with a learning rate of 0.001 and an exponential warmup for the first 50 epochs with no early stopping. 

\subsection{Evaluation metrics}
\label{ssec:evaluation}
We used polyphonic sound event detection scores (PSDS)~\cite{psds} as a performance metric in our studies. The PSDS is more resistant to labeling subjectivity, allowing for ground truth interpretation and temporal structure detection. The single PSDS is computed using polyphonic receiver operating characteristic curves, allowing comparison independent of the operating point. Additionally, it can be adapted for various applications to ensure the appropriate user experience. As a result, it overcomes the limitations of traditional collars-based event F-scores. Using hyperparameters values adopted from the DCASE 2022 Task 4 for the Detection Tolerance Criterion ($\rho_{DTC}$) and Ground Truth intersection Criterion ($\rho_{GTC}$) mentioned in Table~\ref{table:psds}, we compute the PSDS on two scenarios that stress distinct system features. The system must react fast to event detection in Scenario-1, hence it focuses on sound event temporal localization. Scenario-2 focuses less on reaction time and more on class confusion.

\begin{table}[t]
\begin{center}
\vspace*{-\baselineskip}
\caption{PSDS hyperparameters for each evaluation scenario.}
\vspace{-2mm}
\label{table:psds}
\resizebox{0.4\columnwidth}{!}{%
\begin{tabular}{|c|c|c|}
\hline
\textbf{Scenarios} & \hfil \textbf{$\rho_{DTC}$} & \hfil \textbf{$\rho_{GTC}$} \\
\hline
\hline
 Scenario-1     & \hfil 0.7 & \hfil 0.7\\
\hline
 Scenario-2        & \hfil 0.1 & \hfil 0.1\\
\hline
\end{tabular}
}
\end{center}
\vspace{-6mm}
\end{table}

\subsection{Developed system}
The models, DA methods, and experimental settings of our two-stage system are given in Table~\ref{table:sys}. In our two-stage study, Stage-1 uses PANNs while Stage-2 uses FDY-CRNN.

\begin{table}[t!]
\begin{center}
\caption{Description of the two-stage system developed in this work.}
\label{table:sys}
\vspace{-2mm}
\begin{tabular}{|p{0.09\columnwidth}|>{\raggedright}p{0.2\columnwidth}|p{0.5\columnwidth}|}
\hline
\hfil \scriptsize{\textbf{Stage}} & \hfil \scriptsize{\textbf{DA}} & \hfil \scriptsize{\textbf{Description}}\\
\hline\hline
\hfil \scriptsize{1} & \scriptsize{time-masking, frame-shifting, mixup, and Gaussian noise addition} & \scriptsize{We used the architecture given in Section~\ref{ssec:stage1}, trained on weak, unlabeled, and weakified set using exponential softmax function during inference to get the best results.}\\
\hline
\hfil \scriptsize{2} &  \scriptsize{time-masking, frame-shifting, mixup and filter augmentation} & \scriptsize{Stage-1 inferred on the unlabeled set, while Stage-2 used AFL function with $\gamma$=0.625 and $\zeta$=1 to train the architecture specified in Section~\ref{ssec:stage2} on weak, pseudo-weak, and strong sets.}\\
\hline
\end{tabular}
\end{center}
\vspace{-7mm}
\end{table}

\section{RESULTS AND ANALYSIS}
\label{sec:results}
In this section, we report the studies of the proposed two-stage framework in a stage-wise manner, with ablation studies.  

\subsection{Stage-1 comparison}
\label{ssec:stage1}
We are first interested in assessing the contribution of each component to our system at Stage-1. Table~\ref{table:stage1b} shows the SED performance of Stage-1 trained on a real strong set, synthetic strong set, weak set, unlabeled set, and using CNN-14-based PANNs as the pre-trained embeddings. Experimental results show that pre-trained models as feature extractors trained on larger datasets exceed the DCASE 2022 Task 4 organizers' baseline (Baseline), which uses an external dataset. We also observe that our weak training with the PANNs method significantly improves PSDS2 from 0.552 to 0.831 compared to the baseline and drastically decreases PSDS1 from 0.351 to 0.057. PSDS2 increases due to low tolerance in $\rho_{DTC}$ and $\rho_{GTC}$~\cite{psds}, as seen in Table~\ref{table:psds}. As per the parameters for PSDS2, the tolerance value is 0.1, thus the prediction is regarded as true positive even when there is at least one ground truth greater than 1 second out of the 10 seconds clip~\cite{filteraug}. Thus, having a higher PSDS2 is equivalent to having a better AT system. 

This relation is extended to train Stage-2 using weak pseudo-labels from PANNs-based Stage-1 with a higher PSDS2. Further, the decrease in PSDS1 can be attributed to it specifically focusing on temporal localization with a tolerance value of 0.7. Using a pre-trained model also sped up training because the model converged faster with optimized weights. The results show that DA approaches mentioned in Table~\ref{table:sys} and adaptive post-processing improve performance slightly. We also demonstrate the performance of our Stage-1 on the public evaluation set later in Table~\ref{table:publiceval}.

\begin{table}[t]
\begin{center}
\vspace*{-\baselineskip}
\caption{Performance of the baseline and our Stage-1 system on the real validation set of DCASE 2022 Task 4.}
\vspace{-2mm}
\label{table:stage1b}
\resizebox{0.8\columnwidth}{!}{%
\begin{tabular}{|c|l|c|c|}
\hline
\hfil \textbf{Model} & \hfil \textbf{Method} & \hfil \textbf{PSDS1} & \hfil \textbf{PSDS2}\\
\hline
\hline
\hfil CRNN      & DCASE 2022 Task 4 Baseline  & \hfil0.351 & \hfil0.552\\
\hline
\hfil PANNs     & CRNN replaced by CNN-14  PANNs & \hfil0.450 & \hfil0.716\\
\hfil PANNs     & + Weak Training & \hfil0.057 & \hfil0.831\\
\hfil PANNs     & + ICT & \hfil0.067 & \hfil0.834\\
\hfil PANNs     & + DA + Post-processing & \hfil0.075 & \hfil \textbf{0.840}\\
\hline
\end{tabular}
}
\end{center}
\vspace{-3mm}
\end{table}

\subsection{Stage-2 comparison}
\label{ssec:stage2}
To assess the two-stage framework's importance, we constructed the baseline (CRNN) system from the organizers in a two-stage setup. Table~\ref{table:stage2} demonstrates a 5.8\% improvement in total PSDS (PSDS1 + PSDS2) over its result without a two-stage setup in Table~\ref{table:stage1b}. We then used the best Stage-1 CNN-14-based PANNs model (PSDS2 = 0.840) to infer on the unlabeled data to create pseudo-weak labels for Stage-2. Using CRNN in Stage-2, resulted in a PSDS1 of 0.437 and PSDS2 of 0.681. Replacing with FDY-convolutions in Stage-2 improved the SED performance, resulting in a PSDS1 of 0.450, and with additional methods resulted in a PSDS1 of 0.472, a 45.5\% improvement in overall PSDS (PSDS1 + PSDS2) for the two-stage setup. Table~\ref{table:comparisiontable} shows the comparison of our system with other single systems (without ensembling) on the real validation set. The same Stage-2's PSDS1 was 0.479 and PSDS2 was 0.733 on the public evaluation set, as shown in Table~\ref{table:publiceval}.

\begin{table}[t]
\begin{center}
\vspace*{-\baselineskip}
\caption{Performance of the baseline and our Stage-2 system on the real validation set of DCASE 2022 Task 4.}
\vspace{-2mm}
\label{table:stage2}
\resizebox{0.9\columnwidth}{!}{%
\begin{tabular}{|c|l|c|c|}
\hline
\hfil \textbf{Stage-1} & \hfil \textbf{Stage-2} & \hfil \textbf{PSDS1} & \hfil \textbf{PSDS2}\\
\hline\hline
\hfil CRNN      & CRNN  & \hfil0.378 & \hfil0.578\\
\hline
\hfil PANNs      & CRNN  & \hfil0.437 & \hfil0.681\\
\hline
\hfil PANNs      & FDY-CRNN  & \hfil0.450 & \hfil0.701\\
\hfil PANNs      & FDY-CRNN + DA  & \hfil0.468 & \hfil0.714\\
\hfil PANNs      & FDY-CRNN + DA + Post-processing & \hfil0.470 & \hfil0.718\\
\hfil PANNs  & FDY-CRNN + DA + Post-processing + AFL & \hfil \textbf{0.472} & \hfil0.721\\
\hline
\end{tabular}
}
\end{center}
\vspace{-8mm}
\end{table}

\begin{table}[h]
\begin{center}
\vspace*{-\baselineskip}
\caption{Performance comparison of single systems on the real validation set of DCASE 2022 Task 4 with other teams}
\vspace{-2mm}
\label{table:comparisiontable}
\resizebox{0.9\columnwidth}{!}{%
\begin{tabular}{|l|c|c|c|c|}
\hline
\hfil \textbf{Team} & \hfil \textbf{\#Parameters} & \hfil \textbf{PSDS1} & \hfil \textbf{PSDS2} & \hfil \textbf{PSDS1 + PSDS2}\\
\hline
\hline
Ebbers-UPB-task4      & 779M  & \hfil0.505 & \hfil0.807 & \hfil1.312\\
\hline
\textbf{Ours}     & 2.8M & \hfil0.472 & \hfil0.721 & \hfil \textbf{1.193}\\
\hline
 Zhang-UCAS-task4     & 11M & \hfil0.459 & \hfil0.672& \hfil1.131\\
\hline
Kim-GIST-task4     & 1M & \hfil 0.455 & \hfil0.670& \hfil1.125\\
\hline
Dinkel-XiaoRice-task4     & 37M & \hfil0.425 & \hfil 0.644& \hfil1.069\\
\hline
\end{tabular}
}
\end{center}
\vspace{-7mm}
\end{table}

\begin{table}[h]
\begin{center}
\vspace*{-\baselineskip}
\caption{Performance on public evaluation set of DCASE 2022 Task 4.}
\vspace{-2mm}
\label{table:publiceval}
\resizebox{0.6\columnwidth}{!}{%
\begin{tabular}{|c|c|c|c|}
\hline
\hfil \textbf{Stage-1} & \textbf{Stage-2} & \hfil \textbf{PSDS1} & \hfil \textbf{PSDS2}\\
\hline\hline
\hfil Baseline & - &  \hfil0.387 & \hfil0.592\\
\hline
\hfil PANNs & - &  \hfil0.087 & \hfil\textbf{0.801}\\
\hline
\hfil PANNs  & FDY-CRNN &  \hfil \textbf{0.479} & \hfil0.733\\
\hline
\end{tabular}
}
\end{center}
\vspace{-7mm}
\end{table}

\section{CONCLUSION}
\label{sec:conclusion}
In this work, we proposed an AT-assisted sound event detection system using a two-stage framework. We introduced a weak training method to derive weakified labels from strong labels for AT system at Stage-1 and used FDY convolutions in the baseline to focus on the frequency-dependent patterns. Additionally, CNN-14-based pre-trained audio neural networks were used as pre-trained embeddings in Stage-1 to generate reliable pseudo-weak labels to utilize in Stage-2. The studies on DCASE 2022 Task 4 validation set and public evaluation set proved the importance of the proposed two-stage setup and the usage of a weak training strategy. We outperform the DCASE 2022 baseline by 45.5\% on the real validation set in both aspects of the PSDS metric. 

\balance
\bibliographystyle{IEEEbib}
\bibliography{main}

\end{document}